\documentclass[aps, prb, twocolumn]{revtex4}  % for review and submission
 
\usepackage{graphicx}  % needed for figures
\usepackage{bm}        % for math
\usepackage{amssymb}   % for math
\usepackage{amsmath}
\usepackage{epstopdf}

\begin{document}

% The following information is for internal review, please remove them for submission
%\widetext
% the following line is for submission, including submission to the arXiv!!
%\hspace{5.2in} \mbox{Fermilab-Pub-04/xxx-E}

\title{ Integrated optical addressing of an ion qubit }
\author{Karan K. Mehta${^1}$}
%\email{karanm@mit.edu}
\author{Colin D. Bruzewicz${^2}$}
\author{Robert McConnell${^2}$} 
\author{Rajeev J. Ram${^1}$}
\author{Jeremy M. Sage${^2}$}
\author{John Chiaverini${^2}$}

\address{${^1}$Department of Electrical Engineering \& Computer Science and Research Laboratory of Electronics, Massachusetts Institute of Technology, Cambridge, Massachusetts, 02139, USA\\
${^2}$Lincoln Laboratory, Massachusetts Institute of Technology, Lexington, Massachusetts, 02420, USA }
\date{\today}

\maketitle

{\bf The long coherence times and strong Coulomb interactions afforded by trapped ion qubits have enabled realizations of the necessary primitives for quantum information processing (QIP) \cite{haffner2008quantum}, and indeed the highest-fidelity quantum operations in any qubit to date \cite{harty2014high, ballance2015laser, gaebler2016high}. But while light delivery to each individual ion in a system is essential for general quantum manipulations and readout, experiments so far have employed optical systems cumbersome to scale to even a few tens of qubits \cite{monroe2013scaling}. Here we demonstrate lithographically defined nanophotonic waveguide devices for light routing and ion addressing fully integrated within a surface-electrode ion trap chip \cite{chiaverini2005surface}. Ion qubits are addressed at multiple locations via focusing grating couplers emitting through openings in the trap electrodes to ions trapped 50 $\mu$m above the chip; using this light we perform quantum coherent operations on the optical qubit transition in individual $^{88}$Sr$^+$ ions. The grating focuses the beam to a diffraction-limited spot near the ion position with a 2 $\mu$m 1/$e^2$-radius along the trap axis, and we measure crosstalk errors between $10^{-2}$ and $4\times10^{-4}$ at distances 7.5-15 $\mu$m from the beam center. Owing to the scalability of the planar fabrication employed, together with the tight focusing and stable alignment afforded by optics integration within the trap chip, this approach presents a path to creating the optical systems required for large-scale trapped-ion QIP.  }

Individual trapped ions show great promise for quantum computing; however, the lack of a scalable optical interface to manipulate and measure the quantum states of ions has been a major limitation to the development of a large-scale system \cite{monroe2013scaling}. Our approach to this problem utilizes nanophotonic single-mode (SM) waveguides and grating couplers integrated within the trap chip. Light is routed on chip by the waveguides and coupled by the gratings to beams with designed amplitude and phase profiles emitting from the chip towards the ions. These gratings are compact compared to the optical fibers and Fresnel lenses (both with cross-sections $\ge$100 $\mu$m in diameter) previously integrated with planar traps for addressing \cite{kim2011surface} and fluorescence collection \cite{vandevender2010efficient, streed2011imaging}, and most importantly the planar fabrication used here to define the optics for both routing and addressing lends itself to intimate integration with the trap electrodes. Furthermore, such waveguide systems have been demonstrated to be scalable to complex geometries of thousands of devices or more \cite{sun2013large}. Though micro-electro-mechanical systems (MEMS) mirrors integrated with traps have been proposed as well \cite{kim2009integrated}, experiments so far have utilized MEMS components external to the vacuum chamber and separate from the chip \cite{crain2014individual}, leaving full integration an essential outstanding challenge.

\begin{figure}[b]
\centerline{\includegraphics[width=.47\textwidth]{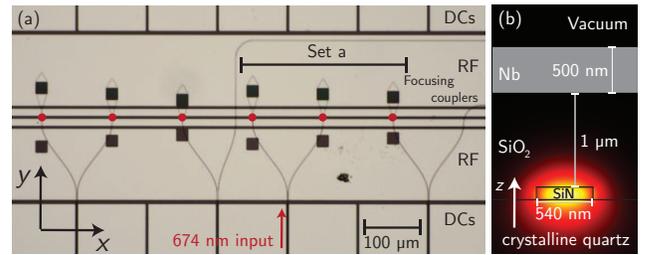}}
\vspace{0 cm}
\caption{\label{fig:schematic} Device layout. (a) Optical micrograph of the designed ion trap with integrated waveguides and couplers underneath at multiple trap zones; waveguides and couplers are visible via topography transfer to the metal. Ions are trapped at one of the positions marked by the red dots, 50 $\mu$m above the electrodes, with appropriate potentials applied to the DC and RF electrodes. (b) Simulated electric field mode profile of the single quasi-TE mode (field oriented predominantly horizontally) waveguide used for routing. The crystalline quartz substrate and PECVD SiO$_2$ form the cladding for the SiN core. }
\end{figure}

\begin{figure*}[t!]
\centerline{\includegraphics[width=1\textwidth]{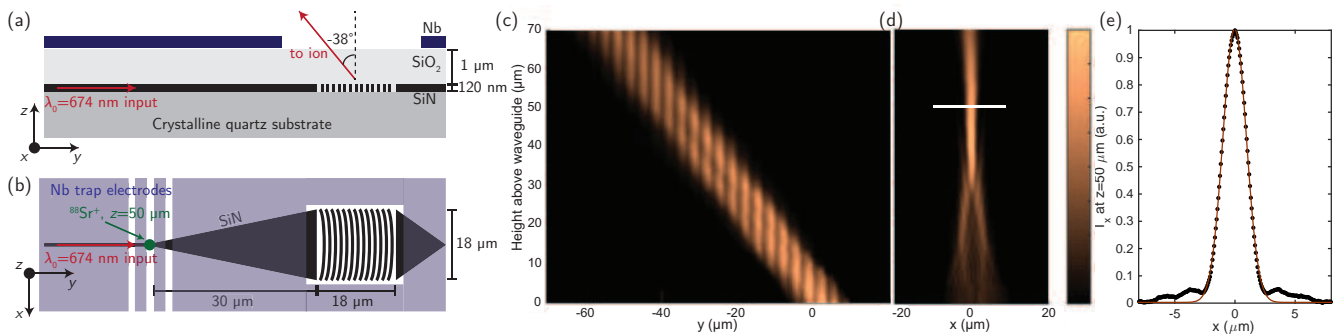}}
\vspace{0 cm}
\caption{\label{fig:focdata} Focusing grating schematic and characterization. (a) and (b) Cross-sectional schematics of the designed focusing grating coupler as integrated with the trap electrodes in the $y-z$ and $x-y$ planes. (c) and (d) ``Knife-edge" profiles of the emitted intensity along $x$ and $y$ from 0-70 $\mu$m above the waveguide layer, showing collimated emission along $y$ and focusing along $x$; striations visible in the $y$-data are interference artifacts owing to reflections in the imaging system. (e) Beam profile along $x$ near the focus, at the designed trap height of about 50 $\mu$m along the section labeled by the white line in (d), the fit (brown line) to the measured points indicates a predominantly Gaussian beam with $1/e^2$ radius $w_0 = 2.0(1)$ $\mu$m, with uncertainty arising from the pixel-length calibration.}
\end{figure*}

Integrated waveguide devices bring several advantages for ion addressing in planar traps. The ability to fabricate, in the same lithographically defined waveguide layer, multiple splitters, waveguide crossings, and bends with radii less than $10$ $\mu$m, would enable the realization of a variety of trapped ion architectures, with flexibility as to arrangement of qubits \cite{kielpinski2002architecture, chiaverini2008laserless}, and with light delivered in parallel to each site. This parallelism will be essential in large-scale systems in which speed is at a premium due to finite coherence times.  Additionally, grating couplers near the ions can focus light to $\mu$m-scale spots, allowing quantum logic gates of a given interaction time using 2-3 orders of magnitude less power when compared to geometries with beams propagating parallel to the chip surface, in which the beam waists are typically limited by diffraction and beam-clipping concerns to $30-50$ $\mu$m diameters \cite{kim2009integrated}. This focusing is crucial also for general individual addressing in an ensemble of closely-spaced ions \cite{schindler2013quantum}. In addition, the phase stability of waveguide approaches even for complex optical paths \cite{politi2008silica} will benefit qubit operations, which are generally phase-sensitive. Furthermore, definition of optics within the trap chip essentially eliminates beam pointing instabilities at the ion location as a noise source \cite{brown2011single, schindler2013quantum}. Beyond trapped-ion QIP, integrated parallel distribution and focusing of light near a chip surface may find further application in atomic physics, such as in ion clocks or neutral atom dipole trap arrays, and more broadly in the various applications of nanophotonic systems.

The trap electrodes and waveguide patterns in the device presented here are visible in the optical micrograph of Fig.~\ref{fig:schematic}(a). Waveguides were fabricated on a crystalline quartz substrate in a silicon nitride (SiN) film (with refractive index $n \approx 2.0$), with cross-sectional areas of approximately 120 nm $\times$ 540 nm, single-moded for the quasi-transverse-electric (quasi-TE) polarization at $\lambda_0 = 674$ nm. These waveguides route light on chip without phase-front distortions or diffraction. Approximately 1 $\mu$m of SiO$_2$ forms the top cladding, above which sit niobium (Nb) trap electrodes; the resulting cross section together with a simulated guided mode E-field profile is illustrated in Fig.~\ref{fig:schematic}(b). This guided mode is coupled to a free-space beam via a focusing grating coupler, which consists first of a taper to expand the mode to that of an 18 $\mu$m-wide waveguide, and then a series of curved grating lines with period, duty cycle, and radius of curvature chosen to couple the light to a beam focused near the ion location and polarized in the $x$ direction (see Methods), illustrated schematically in Fig.~\ref{fig:focdata}(a), (b). 

Light is input to the chip via separate grating couplers designed to couple to a $30$ $\mu$m-diameter beam; light is focused onto these couplers by a 15 cm focal length lens and at an angle $-37^\circ$ from normal to the chip surface. To reduce possible scatter from the input couplers at the ion location, these input couplers are approximately 6.5 mm from the trap center; thus the in-coupled light is routed on chip in a SM waveguide over about 8.5 mm and through two adiabatic 50/50 power splitters to three focusing couplers at the trap site, and to one output waveguide intended to produce an output beam for optimization of the input coupling. The three focusing couplers are offset by different distances from the trap axis to account for possible misalignments between trap sites and beams, as shown in Fig.~\ref{fig:schematic}; two sets of these three were included on the chip (each excited by one input coupler), and trapped-ion measurements were taken with those labeled `Set a' in the micrograph. The emitting region of the coupler has an area of $18 \times 18$ $\mu$m$^2$, and design is summarized in the Methods. 

The ion trap design is as presented previously \cite{sage2012loading}, except for openings in the RF electrodes (introduced symmetrically about the trap axis) to allow the beams from the focusing couplers to emit through the chip surface.  

\begin{figure*}[]
\centerline{\includegraphics[width=.75\textwidth]{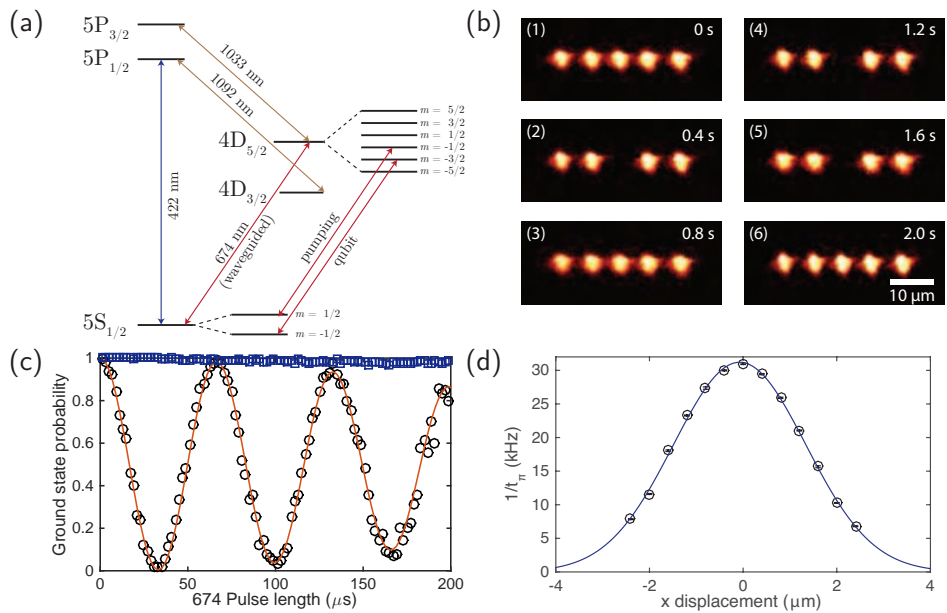}}
\vspace{0 cm}
\caption{\label{fig:rabidata} Addressing and coherent manipulation of individual ions. (a) Relevant level structure of $^{88}$Sr$^+$. (b) Sequence of EMCCD images of 422-nm fluorescence from a chain of 5 ions, with the middle ion aligned to the grating coupler's focus and occasionally entering the dark D state due to the addressing 674 nm beam; the sequence spans 2 seconds with frames evenly spaced. (c) Rabi oscillations on the $5S_{1/2}, m=-1/2 \rightarrow $ $4D_{5/2},m=-3/2$ transition obtained near the focus of the grating coupler. Each point represents the probability that the electron remains in the ground state after a pulse of varying length over 450 repetitions, and the line is a fit to a Rabi oscillation with Gaussian amplitude decay, from which the first $\pi$-rotation's fidelity is determined to be 99\%. Blue squares are with the ion displaced by 7.5 $\mu$m along the trap axis, showing low excitation rate away from the focus. (d) Rabi rates vs. ion position as ion is scanned through the focus along the trap axis, with a Gaussian fit (blue line) with $1/e^2$ half-width of 2.8 $\mu$m, indicating an optical intensity profile with $w_0 = 2.0$ $\mu$m.}
\end{figure*}

The couplers and waveguides were characterized independently of experiments with trapped ions in separate test structures on the same wafer as the trap-integrated devices. The emission from the ion-addressing couplers was characterized by imaging the emitted light at various heights above the waveguides through a microscope with a NA=0.95, 50$\times$ objective. The intensity at different heights was measured by scanning the imaging system away from the waveguide layer with a differential micrometer. The resulting 2D images were integrated along $y$ and $x$ to show beam profiles along $x$ and $y$ respectively, in analogy to a knife-edge measurement. 

Fig.~\ref{fig:focdata}(c) and (d) show the resulting intensity profiles of the emitted light along the $y$ and $x$ directions, showing a collimated beam emerging along $y$, and focusing along $x$ primarily to a spot with a diffraction-limited minimum 1/$e^2$-radius of  $w_0 = 1.8 \pm 0.1$ $\mu$m at 42 $\mu$m, and a slightly expanded waist of $2.0 \pm 0.1$ $\mu$m at the 50 $\mu$m trap height (profile shown in Fig.~\ref{fig:focdata}(e). As the discrepancy in the actual $z$-position of the focus with respect to the target is less than the Rayleigh range along this dimension, the effect on beam waist is small. The simulated efficiency of these couplers is 32\% (see Methods). 

The full ion-trap device was tested in a cryogenic vacuum setup similar to one described previously \cite{sage2012loading}, with the chip at approximately 4K; after loading, ions could be trapped in the present system for over 6 hours with Doppler cooling. A magnetic field of about 6 G was applied perpendicular to the trap surface along $z$ to break the degeneracy of the Zeeman sublevels; the relevant levels are illustrated in Fig.~\ref{fig:rabidata}(a).

Coherent operations here utilized the $\Delta m=-1$ transition. The ion is optically pumped into the  $5S_{1/2}, m=-1/2$ state with six $50$ $\mu$s-long pulses emitted from the focusers (at $\lambda=674$ nm), each followed by quench pulses at 1033 nm (see Fig.~\ref{fig:rabidata}a); the probability that the electron remains in the $S$ orbital is measured by the presence or absence of scattered light when the ion is illuminated with light near resonant with the $5S_{1/2} \rightarrow 5P_{1/2}$ transition at 422 nm, with 1092 nm light also incident during readout to repump out of the $4D_{3/2}$ state. As labeled in Fig~\ref{fig:rabidata}(a), the qubit and pumping frequencies were routed to the ions via the integrated waveguides and couplers; in this work the other wavelengths present were in free-space beams. 

Fig.~\ref{fig:rabidata}(c) shows the probability that a single  ion remains in the ground state after a 674 nm pulse of varying length resonant with the $\Delta m = -1$ transition, with each point representing the average probability inferred from 450 repetitions. With the ion near the beam center, Rabi oscillations with $t_\pi = 33.2$ $\mu$s are observed (black circles), and with the ion displaced by 7.5 $\mu$m, low probability of excitation is observed (blue squares). The ions in this experiment were not cooled to the motional ground state, and thermal occupancy of motional modes contributes to decay in Rabi contrast with increasing pulse length; nevertheless the fidelity of the first $\pi$-rotation is 99\%. We verified also that Rabi oscillations with comparable $\pi$-times could be observed with light from the couplers at all three trap zones in Set a (Fig.~\ref{fig:schematic}), illuminated through cascaded 50:50 splitters from a single waveguide. 

The profile of the beam emitted from the focuser addressing the ion was measured by translating the ion along the trap axis ($x$), and measuring the Rabi oscillation $\pi$-times at various displacements; since the Rabi rate $\Omega_r \propto 1/t_\pi \propto \sqrt{I}$, with $I$ the optical intensity, this corresponds to a measurement of the beam profile along this direction. The points in Fig.~\ref{fig:rabidata}(d) are well fit by a Gaussian (blue line), indicating an intensity profile with $w_0 = 2.0$ $\mu$m. This verifies that the light reaching the ion is predominantly in the focused beam designed.  

That this beam could individually address ions was qualitatively observed with 5 ions trapped in the same well. Quantum jumps between bright and dark states, observed by imaging the chain onto an electron multiplying charged coupled device (EMCCD) camera, occurred only in the center ion aligned to the focus of the center coupler in Set a (Fig.~\ref{fig:schematic}). This is illustrated in the sequence of images in Fig.~\ref{fig:rabidata}(b), spanning 2 seconds, with the inner 3 ions each separated by about 7 $\mu$m. 

\begin{figure}[]
\centerline{\includegraphics[width=.5\textwidth]{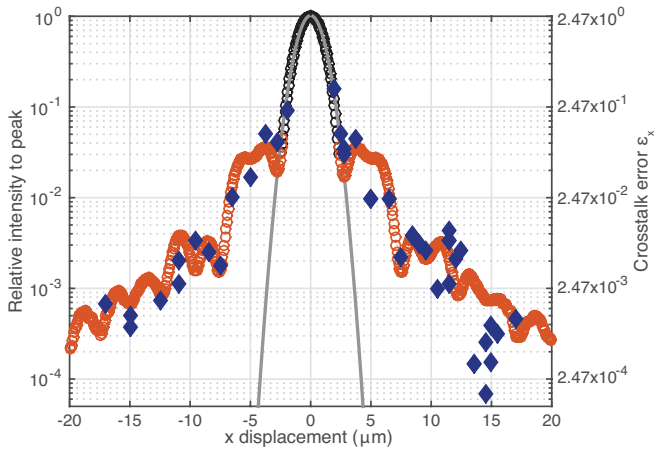}}
\vspace{0 cm}
\caption{\label{fig:xtalkdata} Crosstalk error characterization. Black and red points: imaged intensity of coupler emission along $y$ (see text for description of measurement); and blue diamonds: intensity relative to beam center inferred from cross-talk errors measured with ions variously displaced along the trap axis (with corresponding values of $\epsilon_x$ as defined in the text labeled on the right vertical axis), demonstrating crosstalk errors of order $10^{-3}-10^{-2}$ and below beyond $\pm7.5$ $\mu$m from center, and of order $10^{-4}$ past $\pm12.5$ $\mu$m.}
\end{figure}

Particularly for individual addressing in linear ion chains, crosstalk  between neighboring ions is an important potential error source \cite{knoernschild2010independent, crain2014individual, schindler2013quantum}, and the simple individual addressing afforded by the ability to tightly focus short wavelength radiation \cite{naegerl1999laser} is a significant advantage of optical in relation to microwave approaches \cite{warring2013individual}. We quantified crosstalk errors that would result on a neighboring ion by displacing an ion by a known distance from the focus, and measuring the probability of excitation when a pulse of length equal to the $\pi$-time at the focus, $t_{\pi0}$, is applied; this we define as the crosstalk error, consistent with previous work~\cite{warring2013individual}. This probability of excitation is $\sin^2(\Omega_d t_{\pi0}/2)$, with $\Omega_d$ the Rabi rate at the displaced position; for $\Omega_d t_{\pi0} <<1$, this probability and hence crosstalk error $\epsilon_\times$ is proportional to the ratio of the beam intensity at the displaced position $I_d$ to that at center $I_0$, where $\epsilon_\times = (\pi/2)^2 I_d/I_0$. 

Both the measured intensity profile near the beam center and along the trap axis, measured with a CCD in the imaging setup used for the data in Fig.~\ref{fig:focdata}, as well as values of $I_d$ inferred from $\epsilon_\times$ measured with the ion (blue diamonds), are plotted together in Fig.~\ref{fig:xtalkdata}, showing good correspondence. To obtain sufficient dynamic range, the intensity measurements are taken at two exposure times, with dark frames subtracted; points from the short and long exposure times are shown in black and red, respectively.  Excitation of higher-order spatial modes in the grating region contributes significantly to the observed deviation from the Gaussian profile (gray line) beyond about 2.5 $\mu$m. Optimization of this taper to tailor the transverse field profile in the grating should reduce the crosstalk errors at displacements of around 3-5 $\mu$m, a range typical for many ion trap experiments. For the $^{88}$Sr$^+$ ions used here, at a 1 MHz axial trap frequency the two-ion inter-ion spacing is 4.3 $\mu$m. Relative intensities of ${\sim}1$\% at 5 $\mu$m-displacements, as achieved with the present device, are already comparable to those in existing experiments with individual addressing \cite{schindler2013quantum}; but as operation fidelities increase it will be desirable to reduce crosstalk at such distances to the $10^{-3} - 10^{-4}$ level or below so it is not a dominant noise source. While we expect this to be a nontrivial challenge, engineering the taper from the SM waveguide to the grating offers a route to controlling the field profile along $x$ that may allow similar devices to approach such levels. Recent devices have shown improvement in this metric \cite{mehta2016precise}. 

This work demonstrates the possibility for large-scale nanophotonics integration within planar ion traps, allowing focusing at ion sites, flexible and parallel routing in complex geometries, robust alignment stability, and utilizing scalable planar fabrication. We anticipate a number of avenues for future work. The use of lithographically defined nanophotonic waveguide devices is readily extensible to more complex optical arrangements, at the other wavelengths required for full ion control as well, and co-design with more complex traps; the CMOS-compatibility of both the planar ion traps \cite{mehta2014ion} as well as the optics \cite{orcutt2012open} furthermore suggests an approach for fabrication of eventual large-scale systems. Fiber coupling directly to the trap chip in a scalable fashion, either still with grating couplers \cite{taillaert2002out}, or direct fiber-waveguide interfaces \cite{groblacher2013highly}, presents a significant technical challenge, but would be required both for a scalable, parallel optical interface to the chip, and to realize the full benefit of beam-pointing stability with this approach. Integration of electro-optic modulators \cite{xiong2012low}, or controlled ion movement through the beams formed by the grating couplers \cite{leibfried2007transport, 2016PhRvL.116h0502D}, with the waveguides and gratings here would further enable parallel encoding of quantum operations to multiple sites, in addition to addressing. Given the demonstrated practicality of scaling such planar photonic systems, their ability to operate across a wide range of wavelengths, and the robust alignment and focusing allowed by integration within the trap chip, we expect this approach will significantly reduce the complexity of optical systems required to implement nontrivial QIP with large ensembles of trapped ions. 

\appendix*
\section{Methods Summary}
\subsection{Device design}
Here we summarize the design of the focusing grating couplers addressing the ions in this work. Following Fig.~\ref{fig:focdata}(a) and (b), a taper first expands the mode of the SM waveguide to a larger size laterally; the taper is nonadiabatic and results in curved phase fronts with radius of curvature at the end of the taper approximately equal to the taper length. Subsequently a grating consisting of a series of lines approximately along $x$ with spatially varying period $\Lambda(y)$ and grating strength $\alpha(y)$ emits the light at an angle approximately $-38^\circ$ from normal; backwards emission is preferred to prevent emission into multiple diffraction orders. The lines are parabolic, with curvature radius chosen to focus the beam emitted in free space along the direction transverse to propagation (along $x$ in Fig.~\ref{fig:focdata}), accounting also for the divergence introduced by the non-adiabatic taper. Due to the orientation of the couplers with respect to the trap axis, and that multiple ions in a given trap zone arrange themselves along the trap axis and can be repositioned with DC fields across many microns along the axis, the couplers were designed to focus only along $x$; this eases requirements on alignment between the waveguide and trap metal features. 

The grating ``strength" $\alpha$, defined such that in a uniform periodic structure the intensity guided along propagation direction $y$ would decay as $I \propto e^{-2\alpha y}$, governs the rate of emission of the input light over the course of the grating. In the weak-grating limit of low index contrast, $\Lambda$ would set the emission angle and the grating's duty cycle (defined as the fraction of the local grating period occupied by the SiO$_2$ gaps between grating lines) along $y$ would set $\alpha$, with a higher duty cycle corresponding to a stronger grating and higher $\alpha$; however in the case of higher index-contrast gratings as used here the two are coupled. Hence in the present devices, the periodicity (ranging from 290 - 310 nm) and duty cycle (from 0.1 to 0.4) are varied over the course of the grating along $y$ so as to produce an approximately Gaussian amplitude profile along $y$, while maintaining a constant angle of emission; finite-element simulations of long periodic gratings are used to determine the emission angle and grating strength of a given period and duty cycle, and from these the desired period and duty cycle along $y$ in the focusing gratings are assembled to produce the desired Gaussian amplitude profile, without numerical optimization. The curvature of the grating lines was chosen to produce a beam focused along $x$ near 50 $\mu$m above the waveguides. 

Although the couplers were present on only one side of the trap axis, the 20 $\mu$m-square openings in the RF electrodes of the ion trap (Fig.~\ref{fig:schematic}) were introduced symmetrically around the trap axis to prevent walk off along $y$ of the trapping pseudopotential minimum; along $z$, 3D simulations indicate that these openings resulted in the RF null moving up away from the electrode from the initial $50$ $\mu$m by only 1 $\mu$m. 

The field of the beam is expected to be polarized predominantly along the trap axis. Due to the symmetry of the structure and the SM waveguide feeding the taper and grating (whose dominant $E$-field component along $x$ is even about the $yz$-plane, and whose components along $y$ and $z$ are odd about the $yz$-plane and 0 at the center of the waveguide), at focus the polarization in principle should be purely along $x$, the trap axis. By inserting a polarizer into the microscope used for the beam-profile measurements in Fig.~\ref{fig:focdata}, we measure the power in the transverse field component orthogonal to $x$ over the whole beam to be 2\% of that in the dominant polarization component, and a minimum polarization impurity at center, consistent with the symmetry argument above, of $<5\times10^{-4}$ in relative intensity. Along with these orthogonal transverse components, which result from the taper and grating design, away from the beam center there should additionally be longitudinal field components (pointing along the propagation direction); as these are localized to the focal region, our present polarization measurement is not sensitive to them, though these longitudinal components arise when any Gaussian beam is tightly focused and are not due to the grating design. We leave a full characterization of these field components orthogonal to the dominant one along $x$ and their effects on relative transition rates of the 5$S_{1/2}\rightarrow4D_{5/2}$ sublevels for future work. 

A similar grating was designed for the input coupler, with an emitting region of $40 \times 40$ $\mu$m$^2$, designed to emit (and hence couple to) a free-space beam of approximately $30$ $\mu$m diameter at -$37^\circ$ from normal. Since we presently in-couple from a free-space beam, beam-pointing instability would still appear in coupling variations. Nevertheless, in-coupling goes as the overlap between the approximately Gaussian profile corresponding to the input grating and the Gaussian input beam profile, and since the overlap between two Gaussians of variance $\sigma^2$ with one displaced by $d$ with respect to the other is, as a function of $d$, a Gaussian with variance $2\sigma^2$, for a given $d<\sigma$ the intensity variation at the ion would be ${\sim}2\times$ lower than if the same beam were directly incident on the ion. If furthermore we take into account that the input beam waist $w_g$ when grating-coupled can be chosen to be larger than that of a focused addressing beam $w_i$ directly incident on the ion, supposing comparable beam displacements in both cases, the coupler would offer lower variations by a factor of $w_i^2/2w_g^2$. These approximate considerations indicate that we can expect improvement in pointing stability with the present coupling, though the full benefit of our approach in this regard would be realized with direct fiber-coupling. 

Finally, we note that although light propagating through the trap chip surface may result in photo-induced charging of the dielectric \cite{harlander2010trapped}, we find the compensation voltages are stable over days, suggesting negligible charging from the $674$ nm light; possible charging at shorter wavelengths, if eventually problematic, could be overcome for example by coating the electrode openings with a transparent conductive layer \cite{eltony2013transparent}.

\subsection{Device fabrication}
The 1 cm$^2$ die that formed the chip was written three times on a 3-inch crystalline quartz wafer, chosen as substrate for its high thermal conductivity at low temperature and its relatively low optical index ($n \approx 1.54-1.55$ at $\lambda_0 = 674$ nm) which serves to keep the optical mode well confined in the SiN core. An Oxford-100 plasma-enhanced chemical vapor deposition (PECVD) tool depositing SiN at 300$^\circ$C was used to create the SiN film. Following the HSQ resist spin on and softbake at 85$^\circ$C, to prevent sample charging during e-beam lithography, a thin layer of conductive polymer (E-SPACER 300Z) was spun on top of the HSQ. Electron beam exposure was performed with an 125 keV e-beam lithography system (Elionix F-125). Following exposure the conductive polymer was rinsed off with DI water and the HSQ was developed in a 1\% NaOH, 4\% NaCl solution for 4 minutes, and further rinsed with acetone and isopropyl alcohol. The pattern was transferred to the nitride film via reactive ion etching (RIE) using CHF$_3$ and O$_2$ gases. The same PECVD tool as for the nitride was then used to deposit the SiO$_2$ cladding. Alignment marks written in the nitride were used to spatially reference the photomask for Nb during contact lithography after sputter deposition of the metal film. Following the trap electrode lithography and RIE of Nb in SF$_6$, the individual die were diced from the wafer (leaving the independent test structures used for the grating measurements in Fig.~\ref{fig:focdata} intact), mounted, and the trap electrodes wirebonded. 

Although electron-beam lithography was used to define the waveguides and gratings in this work, the minimum gap size in the grating design here is 30 nm, within resolution limits of current 14-nm CMOS processes. Furthermore silicon nitride waveguides with losses below 1 dB/cm in the visible have been fabricated pholithographically for some time \cite{daldosso2004comparison}. As such we expect it should be possible to produce the same devices in a CMOS process. 

\subsection{Optical losses}
Using a first-principles calculation of the Rabi frequency \cite{james1998quantum}, the 33 $\mu$s $\pi$-time observed, given the measured beam dimensions from the focusing coupler, is consistent with a power of 300 nW being emitted from the grating coupler, 39 dB lower than the ${\sim}2.6$ mW incident on the input coupler. After accounting for the 6 dB designed intensity reduction owing to the two 50/50 splitters in the optical path, the system losses total 33 dB. A number of sources contribute to this loss. Propagation loss in the waveguides was measured in independent test structures to be 6 dB/cm, dominated by material loss; this waveguide loss in our sample contributes 5 dB over the 0.85 cm over which the light is routed on chip. We note that the deposition here was not optimized for loss, but PECVD SiN has been demonstrated elsewhere with material loss as low as 0.1 dB/cm in the red and $< 1$ dB/cm at as low as 470 nm \cite{gorin2008fabrication}. The coupler's simulated efficiency of 32\% (calculated as the upwards-radiated power divided by the input power in waveguide, from a frequency-domain simulation of the grating) corresponds to a loss of 5 dB, and is due to the approximately vertically symmetric structure of the grating which results in about 50\% of the input power being emitted towards both $\pm z$, reflection off of the oxide-vacuum interface, and the finite length and maximum grating strength in the device (18\% is not emitted by the end of the grating, and the inverse taper at the end of the grating is included to prevent reflection and re-emission). Due to an incomplete etch of the SiN waveguide layer, however, the coupler's loss may be as high as 8-9 dB as fabricated. Waveguide bends in the path are estimated to contribute 3 dB as well due to the incomplete etch. The remaining 16-17 dB is likely due to the input coupler (simulated efficiency 10 dB for a perfect Gaussian mode impinging) and any excess loss from the splitters on chip. 

None of these losses are fundamental, and can be significantly reduced. Waveguide material optimization as mentioned can reduce waveguide loss to a level lower than achievable coupler losses. This, together with optimization of the free-space coupler should bring total loss to about 15 dB (10 dB from input coupler, 5 dB from focusing coupler). With more substantial changes, fiber coupling directly to the chip \cite{taillaert2002out, groblacher2013highly}, should allow improvements of input coupling loss to about 2 dB; and incorporation of a bottom reflecting layer can approximately double the focuser efficiency and reduce focuser loss to about 2 dB; therefore, ultimately we expect the total power efficiency can be increased by almost 30 dB.

\subsection*{Acknowledgements}
We thank Isaac Chuang for initial discussions of the approach; Jie Sun, Amir Atabaki, Amira Eltony and Michael Gutierrez for helpful discussions; the MIT Microsystems Technology Laboratory Staff, and the Nanostructures Lab and Mark Mondol in particular with electron-beam lithography; and Peter Murphy, Jeanne Porter, and Chris Thoummaraj for assistance with ion-trap fabrication and packaging. This work was partially funded by NSF program ECCS-1408495. KK Mehta acknowledges support from a DOE Science Graduate Fellowship and the NSF iQuISE IGERT program. 

This work was sponsored by the Assistant Secretary of Defense for Research and Engineering under Air Force Contract \#FA8721-05-C-0002. Opinions, interpretations, conclusions, and recommendations are those of the authors and are not necessarily endorsed by the United States Government.

\subsection*{Author contributions}
KKM, RJR, JMS and JC designed the experiments. KKM designed, fabricated and tested the waveguide devices with RJRÕs supervision. CDB, RM, JMS and JC designed and constructed the vacuum apparatus and laser system used for the ion trap device characterization, and KKM, JMS, and JC performed the single ion experiments with the device. KKM prepared the manuscript and all authors reviewed it and discussed the results. 

\end{document}